\documentclass[journal]{IEEEtran}

\usepackage{tikz}
\usepackage{standalone}
\usepackage[T1]{fontenc}
\usepackage{amsmath}
\usepackage[cmintegrals]{newtxmath}
\usepackage{pgfplots}
\pgfplotsset{compat=1.12}
\usepackage{algorithm}
\usepackage{hyperref}
\hypersetup{colorlinks=true,urlcolor=blue,citecolor=blue,linkcolor=red}
\usetikzlibrary{plotmarks,matrix,chains,scopes,fit,calc,shapes,positioning,decorations,intersections,arrows,backgrounds,shadows}
\newlength\FigureHeight
\newlength\FigureWidth
\def\calA{\mathcal{A}}

\def\emax{E_{\max}}

\def\kmax{K^\bullet}
\def\nch{N_{\text{ch}}}

\colorlet{shapercolor}{green!20!white}
\colorlet{bshapercolor}{cyan!20!white}
\colorlet{feccolor}{blue!20!white}
\colorlet{mapcolor}{red!20!white}
\colorlet{chancolor}{black!10!white}

\def\lf{\left\lfloor}   
\def\rf{\right\rfloor}

\usepackage{amsthm}
\theoremstyle{definition}

\usepackage[noadjust]{cite}

\begin{document}
\title{On Kurtosis-limited Enumerative Sphere Shaping for Reach Increase in Single-span Systems}%
\author{Yunus Can G\"{u}ltekin, Alex~Alvarado, Olga~Vassilieva, Inwoong~Kim, Paparao Palacharla, Chigo~M.~Okonkwo, Frans~M.~J.~Willems
\thanks{Y.C. G\"{u}ltekin, Alex~Alvarado, Chigo~M.~Okonkwo and F.M.J. Willems are with  Eindhoven University of Technology, 5600 MB Eindhoven, The Netherlands. E-mail: y.c.g.gultekin@tue.nl.}
\thanks{Olga~Vassilieva, Inwoong~Kim and Paparao~Palacharla are with Fujitsu Network Communications, Inc., Richardson, 75082 TX, USA.}
\thanks{The work of Y.C. G\"{u}ltekin and A. Alvarado has received funding from the European Research Council (ERC) under the European Union's Horizon 2020 research and innovation programme (grant agreement No 757791).}
}

\maketitle

\begin{abstract}
 The effect of decreasing the kurtosis of channel inputs is investigated for the first time with an algorithmic shaping implementation.
    No significant gains in decoding performance are observed for multi-span systems, while an increase in reach is obtained for single-span transmission.
\end{abstract}

\IEEEpeerreviewmaketitle

\section{Introduction}
Probabilistic amplitude shaping (PAS) is a coded modulation strategy that increases achievable information rates (AIRs) of communication systems~\cite{BochererSS2015_ProbAmpShap,Buchali2016_RateAdaptReachIncrease}. 
This increase can be explained in two ways for the additive white Gaussian noise (AWGN) channel. From a \textit{distribution} point of view, PAS generates inputs with an approximately capacity-achieving distribution~\cite{SchulteB2016_CCDM,Fehenberger2019_MPDM}. From a \textit{coding} point of view, PAS encodes data bits to signal points more energy-efficiently~\cite{Schulte2019_SMDM,GultekinHKW2019_ESSforShortWlessComm}.

For fiber-optical channels, the enhanced Gaussian noise (EGN) model predicts that inputs with high kurtosis lead to a high nonlinear interference (NLI), and hence, a low effective signal-to-noise ratio (SNR)~\cite{Carena2014_EGNmodel}. 
The kurtosis-dependency of NLI is often used to explain the SNR-penalty observed with AWGN-optimal shaping: Gaussian-like distributions have high kurtosis~\cite{FehenbergerABH2016_OnPSofQAMforNLFC,Renner2017_ExperimentalPScomparison}.

\begin{figure}[t]
    \centering
	\resizebox{\columnwidth}{!}{\includegraphics{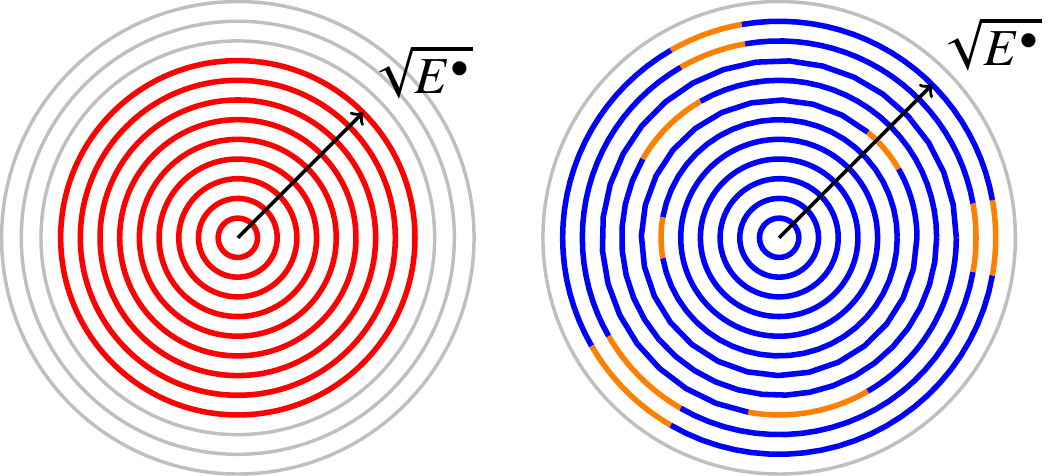}}
    \caption{Shell occupation in an $n$-dimensional sphere:
(Left, red) sphere shaping, (right, blue) kurtosis-limited sphere shaping. Since a higher energy usually implies a higher kurtosis, the sequences that are excluded due to $\kmax$ (orange) are mostly from outermost shells.}
    \label{fig:sphericalillustration}
\end{figure}

Kurtosis has mostly been studied from a distribution but not from a coding point of view: the input was optimized to maximize the SNR or the AIR~\cite{Pan2016_PS16QAMinWDM,FehenbergerABH2016_OnPSofQAMforNLFC,Sillekens2018_nonlinearitytailoredPS,Renner2017_ExperimentalPScomparison}.
In~\cite{Yoshida2020_HiDMmasive}, an amplitude shaper was designed specifically for kurtosis reduction based on lookup tables (LUTs). 
We recently proposed kurtosis-limited sphere shaping (KLSS), which extends the idea of sphere shaping by introducing a constraint on the kurtosis of amplitude sequences, in addition to the constraint on their energy~\cite{gultekin_kess_arxiv}.
Adopting a coding point of view, KLSS creates a spherical-like signal space from which high-kurtosis signal points are excluded, as shown in Fig.~\ref{fig:sphericalillustration}.
The resulting signal structure is more suitable for fiber-optical channels.
Unlike a LUT-based approach, KLSS is implemented with a \textit{constructive} shaping algorithm, kurtosis-limited enumerative sphere shaping (K-ESS). 
To the best of our knowledge, K-ESS\cite{gultekin_kess_arxiv} is the only constructive kurtosis-limited shaping algorithm available in the literature.

This paper complements the introduction of K-ESS by studying its performance for short- and long-haul systems.
The obtained conclusion is that kurtosis plays the most significant role for short distances~\cite{Dar2014_AccumultionNLIN,Pan2016_PS16QAMinWDM}, and thus, K-ESS provides improvements for single-span systems.
In single-span systems, the gains are shown to be larger for systems with fewer wavelength division multiplexing (WDM) channels. 
This conclusion agrees with the results from the literature ~\cite{FehenbergerABH2016_OnPSofQAMforNLFC,Renner2017_ExperimentalPScomparison,Sillekens2018_nonlinearitytailoredPS} predicting that gains obtained by optimizing inputs specifically for optical channels exist, but are limited.
Our main contribution is to investigate the effect of reduced kurtosis and confirm via an actual shaping algorithm that kurtosis-limited probabilistic shaping seems to be beneficial only for single-span systems with a few WDM channels.

\section{Sphere Shaping for Optical Channels}

In PAS, amplitudes of the channel inputs are determined by a shaper.
Sphere shaping is a shaping approach where amplitude sequences from within an $n$-spherical shaping set (see Fig.~\ref{fig:sphericalillustration}~(left))
\begin{equation}
\calA^\bullet \!=\! \left\{a^n = (a_1,a_2,\dotsc,a_n)  : \sum_{i=1}^n a_i^2 \leq \emax \right\}, \label{eq:ElimitedSet}
\end{equation}
are used, where $a_i\in \{1, 3,\dotsc, 2^m-1\}$ for integer $m\geq 1$.
For decreasing values of $\emax$, the shaping rate $k/n = \lf\log_2|\calA^\bullet|\rf/n$~(in bits per amplitude) decreases.
Shaping increases energy-efficiency and leads to Gaussian-like 1D symbol distributions, and thus, increases the AIRs for the AWGN channel~\cite{Gultekin2018_ConstShapforShortBlocks}.
Sphere shaping has also been shown to improve performance for optical links~\cite{Goossens2019_FirstExperimentESS,Amari2019_ESSreachincrease,Amari2019_IntroducingESSoptics,Skvortcov2021_HCSSforExtendedReachSingleSpanLins}.

\begin{figure}[t]
\centering
\resizebox{\columnwidth}{!}{\includegraphics{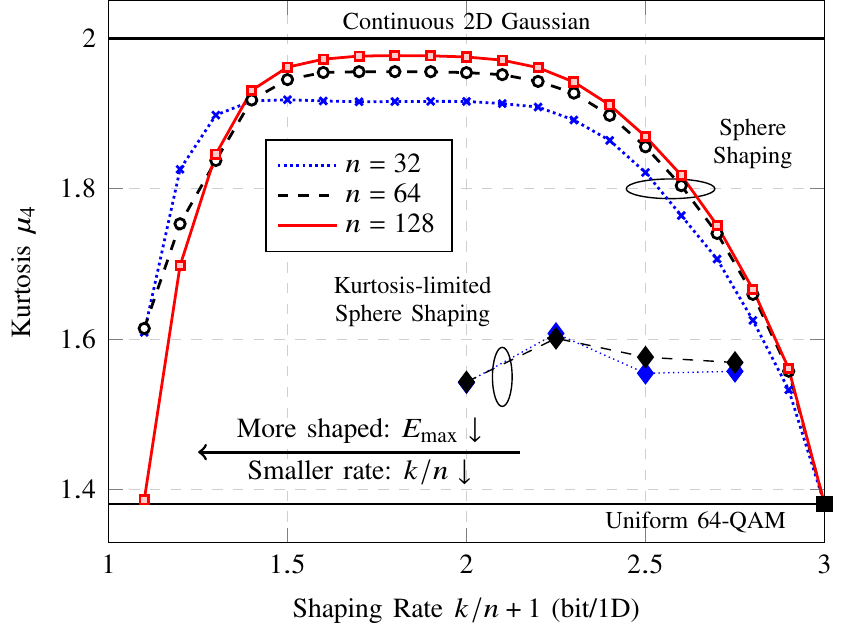}}
\caption{Kurtosis of the channel inputs obtained using sphere shaping with 64-ary quadrature amplitude modulation (64-QAM, $m=3$). 
The blue and the black diamonds correspond to KLSS with $n=32$ and $64$, respectively.
}
\label{fig:EssKurtosis}
\end{figure}

For optical channels, the effective SNR depends on the fourth order standardized moment~\cite{Carena2014_EGNmodel}
\begin{equation}
\mu_4 = \frac{E[|X-E[X]|^4]}{(E[|X-E[X]|^2])^{2}}, \label{eq:highermoementbootom}
\end{equation}
of the channel input $X$.
In Fig.~\ref{fig:EssKurtosis}, $\mu_4$ is plotted for sphere shaping. 
With respect to uniform signaling, sphere shaping has an increased kurtosis which is close to that of 2D Gaussian distribution for shaping rates $0.5\leq k/n \leq 1.5$. 
Furthermore, for increasing values of blocklength $n$, kurtosis increases.
We believe that this might (partly) explain the results observed in the literature, where larger $n$ result in lower effective SNRs~\cite{Amari2019_IntroducingESSoptics,Fehenberger2020_impactofinitelengthshaping,Civelli2020_interplayPSandCPR,Skvortcov2021_HCSSforExtendedReachSingleSpanLins}.

\section{Kurtosis-limited Sphere Shaping}

To devise a shaper that produces inputs with low kurtosis, we define the kurtosis-limited sphere shaping (KLSS) set (see Fig.~\ref{fig:sphericalillustration}~(right))
\begin{equation}
\calA^\blacktriangle \!=\! \left\{a^n  : \sum_{i=1}^n a_i^2 \leq \emax \mbox{ and } \sum_{i=1}^n a_i^4 \leq \kmax  \right\}. \label{eq:EKlimitedSet}
\end{equation}
The shaping rate $\lf\log_2 |\calA^\blacktriangle|\rf/n$ can be adjusted by changing the values of the maximum-energy $\emax$ and maximum-kurtosis $\kmax$ constraints.
As $\kmax$ decreases, sequences with high kurtosis are eliminated from the shaping set, and hence, the rate decreases.
To compensate for this decrease and to keep the rate constant, $\emax$ should be increased (see Fig.~\ref{fig:sphericalillustration}). 
By varying $\emax$ and $\kmax$, a trade-off between energy-efficiency and kurtosis is created.
The sphere shaping set $\calA^\bullet$ \eqref{eq:ElimitedSet} is the same as KLSS set $\calA^\blacktriangle$ \eqref{eq:EKlimitedSet} for $\kmax\rightarrow\infty$.
Thus, KLSS can be seen as a generalization of sphere shaping.

In Fig.~\ref{fig:EssKurtosis}, the minimum $\mu_4$ that can be obtained with KLSS is shown with diamonds.
The minimum is found by computing $\mu_4$ of KLSS for all pairs of $(\emax,\kmax)$ that satisfy the rate constraint according to~\eqref{eq:EKlimitedSet}.
We see that the kurtosis-limitation indeed leads to smaller $\mu_4$ than that of sphere shaping at the same rate.
Furthermore, $\mu_4$ is relatively flat and independent of $n$ for KLSS.
To confirm that smaller-$\mu_4$ inputs obtained with KLSS lead to improved performance, an actual constructive shaping algorithm was proposed~\cite{gultekin_kess_arxiv}.
This shaper is a modified version of the enumerative sphere shaping algorithm~\cite{GultekinWHS2018_ApproxEnumerative} and we call it K-ESS.

\section{Numerical Results}

We simulated optical transmission based on a standard
single-mode fiber (SSMF) with an attenuation of 0.19 dB/km, a dispersion parameter of 17 ps/nm/km, and a nonlinear parameter of 1.3 1/W/km.
At the end of single-span links, there is an erbium-doped fiber amplifier (EDFA) with a noise figure of 5.5 dB.
In the multi-span scenario, spans are of length 80 km, separated by the same EDFA.
The transmitter generates a dual-polarized 56 Gbaud 64-QAM signal (with a root-raised-cosine pulse with 10\% roll-off) using WDM with 62.5 GHz spacing.
For multi-span transmission, the number of channels is $\nch \in\{1, 11, 21, 31\}$.
For single-span transmission, $\nch\in\{1, 11, 41\}$.

The 648-bit low-density parity-check codes of the IEEE 802.11 standard are used for forward error correction (FEC).
The target information rate is $8$~bit/4D (448 Gbit/s) which is achieved using the rate-2/3 code for uniform signaling.
For shaped signaling, PAS is realized with the rate-5/6 code and an amplitude shaper with shaping rate $k/n=1.5$~bit/amplitude.
Both ESS and K-ESS are realized with $n=108$.
All pairs of $(\emax,\kmax)$ that satisfy the rate constraint for KLSS were simulated and the one that leads to the best performance was chosen.
The performance metric is the frame error rate (FER) at the optimum launch power where a frame is a 648-bit FEC codeword.

\begin{figure}[t]
\centering
\resizebox{\columnwidth}{!}{\includegraphics{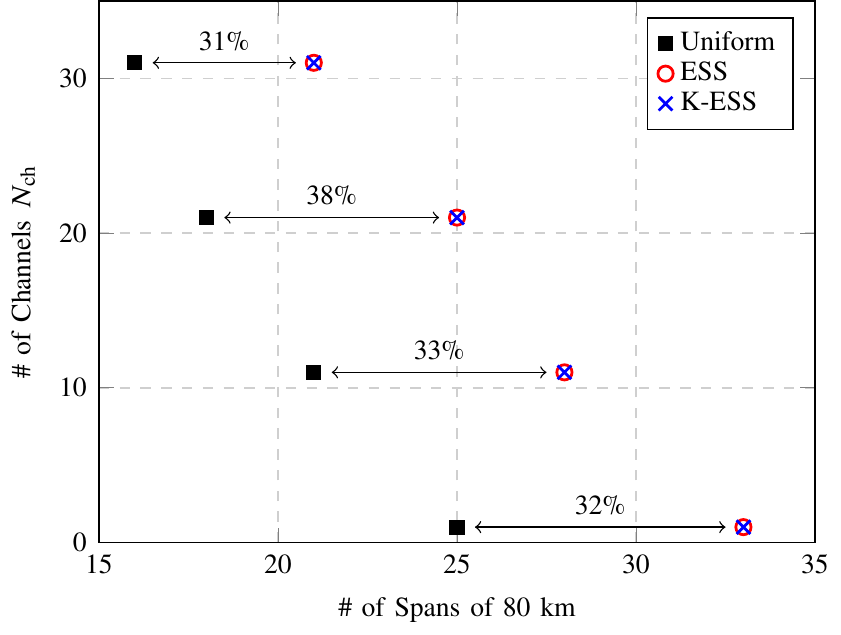}}
    \caption{Reach increase obtained using sphere shaping and kurtosis-limited sphere shaping for long-haul transmission.}
    \label{fig:longhaul}
\end{figure}

In Fig.~\ref{fig:longhaul}, the maximum number of spans at which a FER of $10^{-3}$ can be achieved is shown for multi-span transmission.
We see that more than 30\% reach increase can be obtained for long-haul systems using AWGN-optimal shaping schemes such as ESS. These gains are comparable to the values reported in~\cite{Buchali2016_RateAdaptReachIncrease,Amari2019_IntroducingESSoptics}.
For such systems, the optimum K-ESS scheme is the one that has an inactive kurtosis constraint, which corresponds to ESS.
Thus, for multi-span systems, AWGN-optimal inputs perform well enough that further optimizations based on kurtosis do not provide additional improvement.
This is in agreement with~\cite{FehenbergerABH2016_OnPSofQAMforNLFC} where no improvement in AIR is obtained by optimizing the input based on the EGN model.

\begin{figure}[t]
\centering
\resizebox{\columnwidth}{!}{\includegraphics{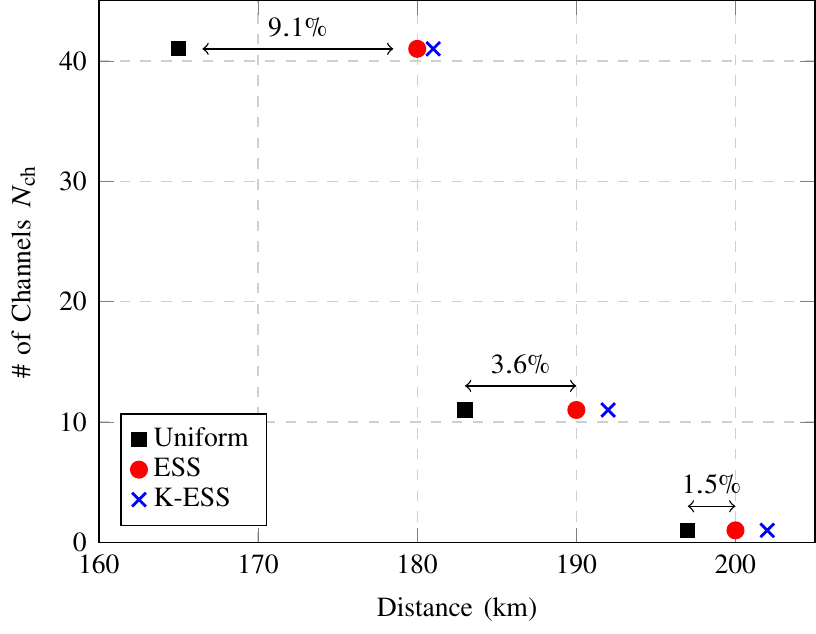}}
    \caption{Reach increase obtained using sphere shaping and kurtosis-limited sphere shaping for single-span transmission.}
    \label{fig:shorthaul}
\end{figure}

In Fig.~\ref{fig:shorthaul}, the maximum distance at which a FER of $10^{-3}$ can be achieved is shown for single-span transmission.
We see that the largest improvement provided by K-ESS is obtained for $\nch=1$.
In this case, ESS provides a 1.5\% reach increase over uniform signaling, while the increase is 2.5\% for K-ESS.
As $\nch$ increases to 11 and then to 41, the reach increase by ESS also increases to 3.6\% and then to 9.1\%.
However, with increasing $\nch$, the performance of K-ESS converges to that of ESS.
These observations imply that for single-span systems, as $\nch$ increases, AWGN-optimal inputs are again good enough, and the importance of kurtosis in the optimization of the input decreases.
This conclusion is in agreement with~\cite{Renner2017_ExperimentalPScomparison} where only marginal improvements in SNR and AIR were obtained by optimizing the input distribution based on the EGN model for single-span transmission with $\nch=9$.

\begin{figure}[t]
\centering
\resizebox{\columnwidth}{!}{\includegraphics{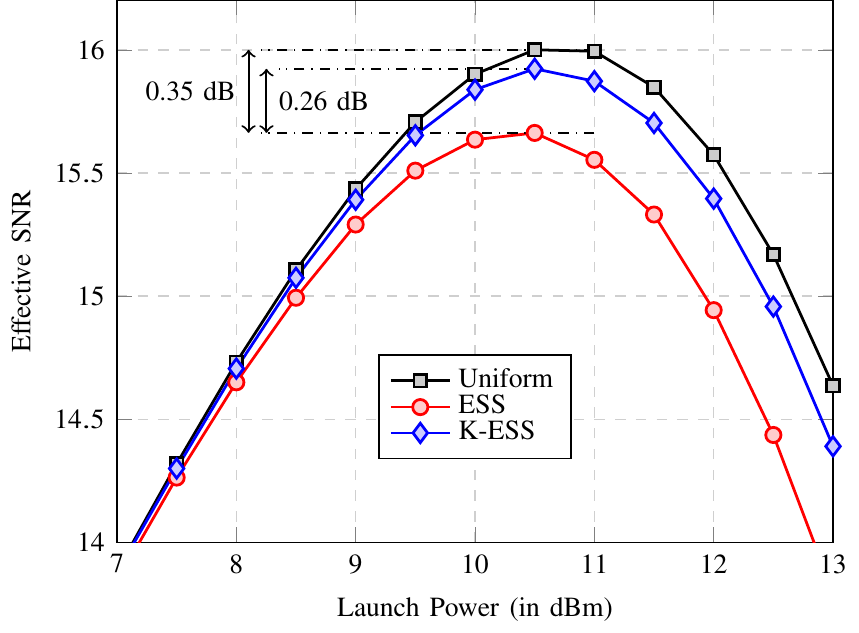}}
    \caption{Launch power vs. effective SNR for single-span single-channel transmission over 197 km of SSMF.}
    \label{fig:effsnr_singlespan}
\end{figure}

In Fig.~\ref{fig:effsnr_singlespan}, effective SNR is shown for single-channel transmission over 197 km SSMF, where uniform signaling achieves a FER of $10^{-3}$.
ESS has a 0.35 dB SNR-penalty with respect to uniform signaling due to its high kurtosis.
K-ESS recovers 0.26 dB of this loss. 
For this scenario, uniform signaling achieves the largest SNR, while providing the worst decoding performance (see Fig.~\ref{fig:shorthaul} at $\nch=1$).
This is due to the shaping gain obtained with ESS and K-ESS.
K-ESS obtains a better trade-off between the shaping gain and the kurtosis, and achieves the smallest FER, or equivalently, the largest reach.

\section{Conclusions}

We have discussed kurtosis-limited sphere shaping (KLSS) as an amplitude shaping approach to generate shaped channel inputs with low-kurtosis.
An algorithmic implementation of KLSS is realized based on enumerative sphere shaping (ESS) and is called K-ESS.
Numerical simulations demonstrate that the reduction in kurtosis and the resulting decrease in the nonlinear interference do not lead to any improvement in performance for long-haul systems.
For single-span systems, however, an increase in reach is obtained, especially for a relatively small number of channels.
Future work should focus on shaping algorithms better tailored to the nonlinear optical channel.

\bibliographystyle{IEEEtran}
\bibliography{IEEEabrv,PhD_refs}

\end{document}